\documentclass[%
 reprint,
 superscriptaddress,
 showkeys,
 amsmath,amssymb,
 aps,
]{revtex4-2}

\usepackage{graphicx}
\usepackage[OliveGreen,dvipsnames]{xcolor}
\usepackage{dcolumn}
\usepackage{bm}

\usepackage{subfiles}

\newcommand{\RK}[1]{{\color{black} #1}}
\newcommand{\JP}[1]{{\color{black}  #1}}
\newcommand{\MB}[1]{{\color{black} #1}}

\newcommand{\C}{{\mathbf C}}
\newcommand{\btau}{\mbox{\boldmath $\tau$}}

\begin{document}

\preprint{APS/123-QED}

\title{Polymeric diffusive instability leading to elastic turbulence in plane Couette flow}

\author{Miguel Beneitez}
\affiliation{
 DAMTP, Centre for Mathematical Sciences, Wilberforce Road, Cambridge CB3 0WA, UK
} 
 
\author{Jacob Page}%
\affiliation{
 School of Mathematics, University of Edinburgh, EH9 3FD, UK
}

\author{Rich R. Kerswell}
\affiliation{
 DAMTP, Centre for Mathematical Sciences, Wilberforce Road, Cambridge CB3 0WA, UK
}

\date{\today}

\begin{abstract}
Elastic turbulence is a chaotic flow state observed in dilute polymer solutions in the absence of inertia. It was discovered experimentally in circular geometries and has long been thought to require a finite amplitude perturbation in parallel flows. 
Here we demonstrate, within the commonly-used \JP{Oldroyd-B and} FENE-P models, that a self-sustaining chaotic state can be initiated via a linear instability in a simple inertialess shear flow caused by the presence of small but non-zero diffusivity of the polymer stress.
Numerical simulations show that the instability leads to a three-dimensional self-sustaining chaotic state, which we believe is the first reported in a \JP{wall-bounded}, parallel, inertialess viscoelastic flow. 
\end{abstract}

\maketitle

Dilute polymer solutions are ubiquitous in everyday life (e.g. foods, shampoo, paints, cosmetics) and understanding how they behave is important for many industrial processes (e.g. plastics, oil, pharmaceuticals and chemicals). The stretching and subsequent relaxation of the polymers introduces new viscoelastic stresses in the flow which depend on the flow’s deformation history. As a result, polymer flows can exhibit startlingly different behaviour from that of a Newtonian fluid like water (\,e.g. rod-climbing, die swell and elastic recoil \cite{Bird_87}\,).  Perhaps most strikingly, a chaotic flow state -- so called ``elastic turbulence'' (ET) -- can occur for vanishing inertia, in stark contrast to Newtonian fluid mechanics where inertia provides the only nonlinearity.
 ET has important applications in small scale flows where, for example, enhanced mixing for chemical reactions or heat transfer for cooling computer chips are highly desirable \cite{groisman2001efficient,squires2005microfluidics}. Even though the origin of this behaviour is still not understood, the key ingredients are believed to be fluid elasticity provided by the polymers and streamline curvature which together give rise to a new elastic linear instability \cite{shaqfeh96, groisman2000elastic,larson2000turbulence,groisman2004elastic}. Experiments in curved geometries confirm the linear instability leads to sustained ET \cite{steinberg2021elastic}.  In contrast, \MB{wall-bounded} inertialess parallel polymer flow has been presumed linearly stable, and experimental work there has focused on triggering a finite amplitude instability instead. Obstacles in the flow have been used to  provide the required streamline curvature believed necessary for this elastic instability and ultimately ET \cite{Pakdel_96, Pan_13, van2022characterizing, datta2022perspectives}. However, the requirements for both initiating such a transition and for the existence of a self-sustaining chaotic state in a planar geometry are unknown. 
 
%
%
Recently, by exploring the large relaxation-time limit of the polymers, a new elastic linear instability has been identified for the parallel flow of an Oldroyd-B fluid in a pipe or channel at finite inertia \cite{garg2018viscoelastic,khalid2021centre} and at vanishing inertia in a channel only 
\cite{khalid2021continuous}.
This instability \JP{is a `centre mode'} -- concentrated \JP{around the centreline of} the channel -- and is strongly subcritical, giving rise to an arrowhead-shaped travelling wave solution \cite{page2020exact} over a large region of the parameter space. This solution has been seen in simulations at both finite \cite{dubief2022first} and vanishing inertia \cite{morozov2022coherent,buza2022finite}, and may play a role in two-dimensional elasto-inertial turbulence where inertia is important \cite{choueiri2021experimental}. \JP{While no link has yet been found between this structure and ET in wall-bounded flows, similar arrowhead-like flow structures have been observed in doubly-periodic `Kolmogorov flow' \citep{berti2008two,berti2010elastic} where a self-sustaining, two-dimensional chaotic state is maintained in the absence of inertia. 
The chaotic dynamics found here can be directly connected to a linear instability of the basic state \citep{boffetta2005viscoelastic,bistagnino2007nonlinear} -- though whether this instability is driven by the same mechanism as the centre mode, or whether the nonlinear chaotic state is a manifestation of the three-dimensional ET found in wall bounded flows is an open problem. 
} \JP{In contrast to these configurations, the simpler case of} constant shear between two differentially-moving, parallel plates, known as plane Couette flow, \JP{has been} considered linearly stable for all inertia and elasticity parameters \cite{garg2018viscoelastic}. 

%
%

In this Letter we report that viscoelastic plane Couette flow is linearly unstable if polymer stress diffusion is included in the model. 
This diffusion is generally so small that it is ignored as an important physical effect, but is reintroduced as a much larger `artificial' diffusion to stabilise time-stepping schemes if their inherent numerical diffusion isn't sufficient.
This diffusion-induced linear instability 
\RK{is distinctly different} from the centre-mode present in channels, being concentrated instead at the walls. Significantly, there is no smallest diffusion threshold below which the instability vanishes: the wavelength of the instability decreases with the size of the diffusion so the instability could be misunderstood as a numerical instability. The growth rate of the instability tends to a non-zero limit as the polymer diffusion goes to zero so the vanishing-diffusion limit is singular. The new diffusive instability exists over a very wide area of the parameter space and is robust to the choice of boundary conditions on the polymer conformation. Direct numerical simulations \MB{(DNS)} show that the instability saturates onto a low-amplitude limit cycle in two-dimensions. Three-dimensional simulations show a transition to sustained spatiotemporal chaos, which we believe to be the first reported computation of such a state in a planar geometry. 

We consider the inertialess flow of an incompressible, viscoelastic fluid between infinite plates at $y=\pm h$ moving with velocity $\pm U_0 \hat{\mathbf x}$. The governing equations are 
\begin{subequations}
\begin{align}
    \boldsymbol \nabla p = \beta \Delta \mathbf u &+ (1-\beta) \boldsymbol \nabla \cdot \mathbf T(\mathbf C), \label{1a}\\
    \boldsymbol \nabla \cdot \mathbf u &=0, \label{1b}\\
        \partial_t \mathbf C + \left( \mathbf u \cdot \boldsymbol \nabla\right) \mathbf C + \mathbf T(\mathbf C) &= \mathbf C \cdot \boldsymbol \nabla \mathbf u +   \left( \boldsymbol \nabla \mathbf u \right)^{T} \cdot \mathbf C + \varepsilon \Delta \mathbf C.  \label{1c}
\end{align}
\label{eqn:governing}
\end{subequations}
where the polymeric stress $\mathbf T$ is related to the conformation tensor $\mathbf C$ using the FENE-P model,  
\begin{equation*}
    \mathbf T(\mathbf C) := \frac{1}{Wi} \left[ \frac{\mathbf C}{(1-(\mathrm{tr}\, {\mathbf C}-3)/L^2_{max}} - \mathbf{I} \right].
\end{equation*}
This model successfully predicted the phenomenon of elasto-inertial turbulence (EIT) in 2010 \cite{Dubief10} which was observed a year later in experiments \cite{Hof11,Samanta13}.

The equations are non-dimensionalised by the half the gap width, $h$, and the plate speed $U_0$, which defines the Weissenberg number $Wi:= \lambda U_0 / h$ (the ratio of the polymer relaxation time $\lambda$ to a flow timescale). The parameter $\beta := \mu_s/\mu_T$ is the ratio of the solvent-to-total viscosities while $\varepsilon:=D/U_0 h$ is the nondimensionalisation of the polymer diffusivity $D$ \cite{Leal_89}.
In this configuration, the laminar basic state is simply ${\mathbf U}= y\hat{\mathbf x}$ with only $T_{xx}=2Wi$ and $T_{xy}=1$ being non-zero stress components for an Oldroyd-B fluid ($L_{max} \rightarrow \infty$).
%
%
The polymer equation (\ref{1c}) changes character from hyperbolic at $\varepsilon=0$ to parabolic for $\varepsilon \neq 0$ and extra boundary conditions are then needed. Three boundary conditions are considered: (i) application of the governing equations with $\varepsilon = 0$ at the walls \cite{sureshkumar1997direct}, (ii) application of the governing equations with only the term $\varepsilon \partial_y^2 C_{ij}$ removed 
\cite{dubief2022first} 
and (iii) Neumann, so $\partial_y C_{ij} = 0$. 

The linear stability of the basic state is examined by introducing small perturbations of the form $ \phi'(\mathbf x, t) = \hat{ \phi}(y)\exp \left(i k_x (x - ct)\right) + c.c.$, where $k_x \in \mathbb R$ is the streamwise wavenumber and $c = c_r + i c_i$ a complex wavespeed, with instability if $c_i  > 0$.  The linear eigenvalue problem is solved by expanding each flow variable using the first $N$ Chebyshev polynomials ($N = 300$ is usually sufficient to ensure convergence). An example eigenvalue spectrum for an Oldroyd-B fluid with $\textit{Wi}=100$, $\varepsilon=10^{-3}$, $\beta=0.95$ and $k_x=3$ is reported in figure \ref{fig1} (the bottom right inset shows the equivalent spectrum with $\varepsilon=0$).
The continuous spectra in the absence of polymeric diffusion are regularised with the introduction of $\varepsilon\neq 0$ \cite{Kumaran_09}, with a pair of linear instabilities emerging with wavespeeds $c_r \sim \pm 1$.
%
%
\begin{figure}
\includegraphics[width=0.95\columnwidth]{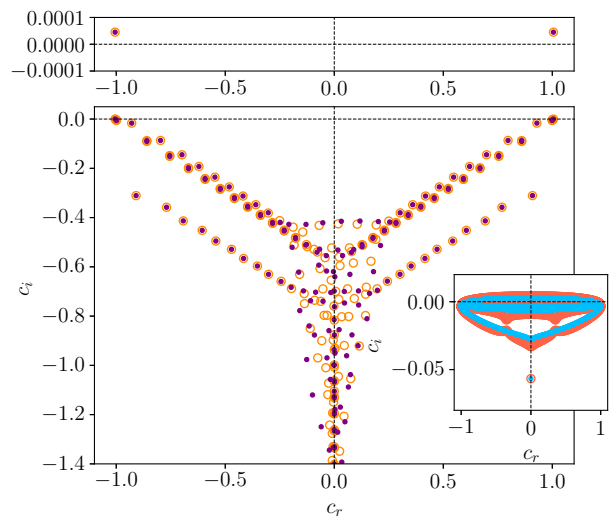}
\caption{\label{fig1}Spectrum at $\textit{Re}=0$, $\textit{Wi}=100$, $\beta=0.95$, $k_x=3$ for two different resolutions $N_y=300$ (orange circles) and $N_y=400$ (purple dots) with $\varepsilon=10^{-3}$ within the domain and $\varepsilon=0$ at the boundaries. Top inset: zoom-in to the unstable eigenvalues $c_i>0$. Bottom right inset: spectrum for the same parameters but $\varepsilon=0$ everywhere in the domain and boundaries and two different resolutions $N_y=300$ (red circles) and $N_y=400$ (blue dots). The most unstable eigenvalues with finite $\varepsilon$ here are
$c = \pm 1.0052607137+4.513293285i\times 10^{-5}$
\JP{Note the apparent instabilities in the continuous spectrum seen in the $\varepsilon=0$ results is associated with poor resolution of the (non-smooth) eigenfunctions in our numerics (e.g. \cite{wilson1999structure}) and can be suppressed if the resolution is increased further.}
} \label{fig:spectra}
\end{figure}
%
%
\begin{figure}
     \includegraphics[width=0.89\columnwidth]{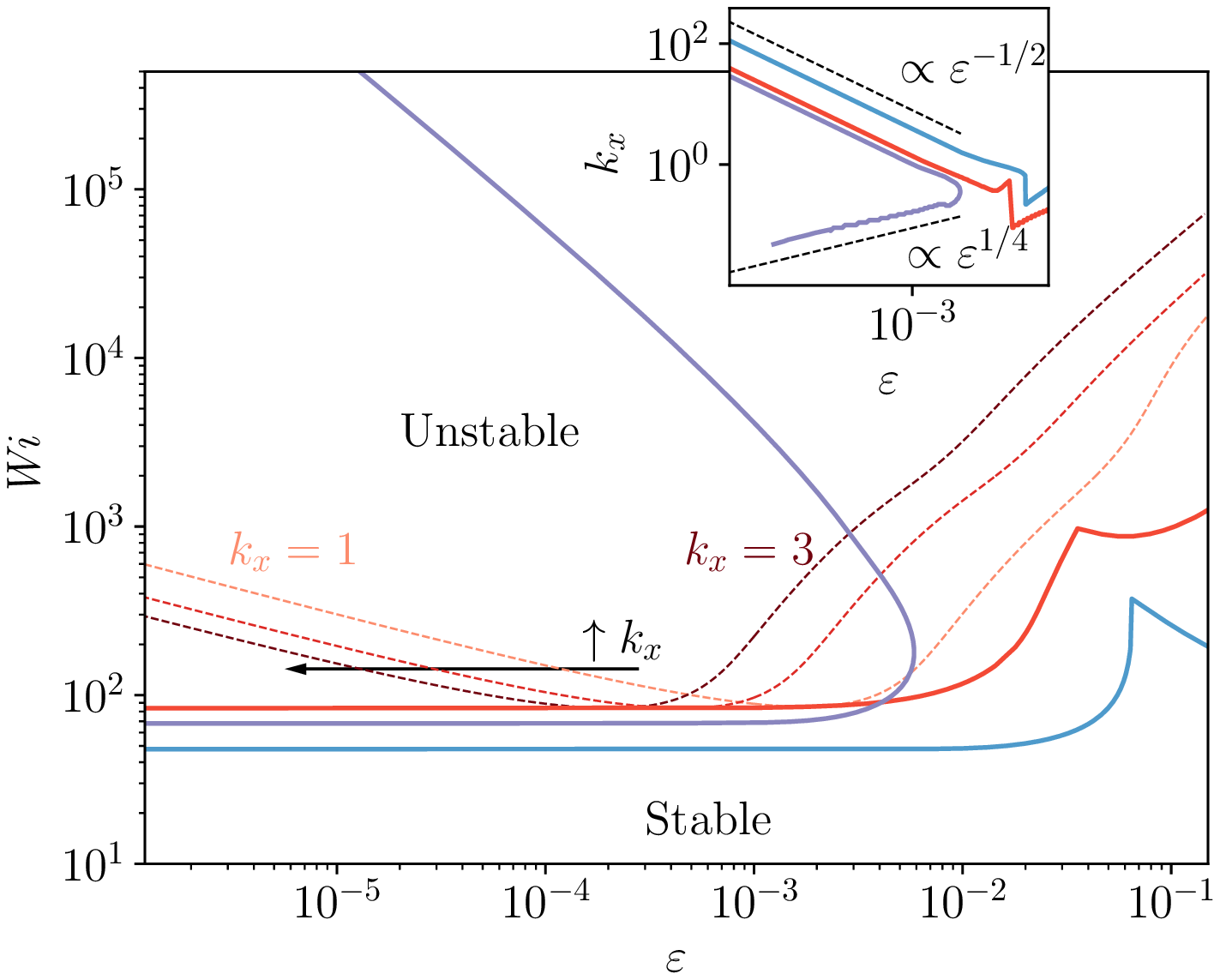}\put(-225,155){$(a)$} \\
    \includegraphics[width=0.89\columnwidth]{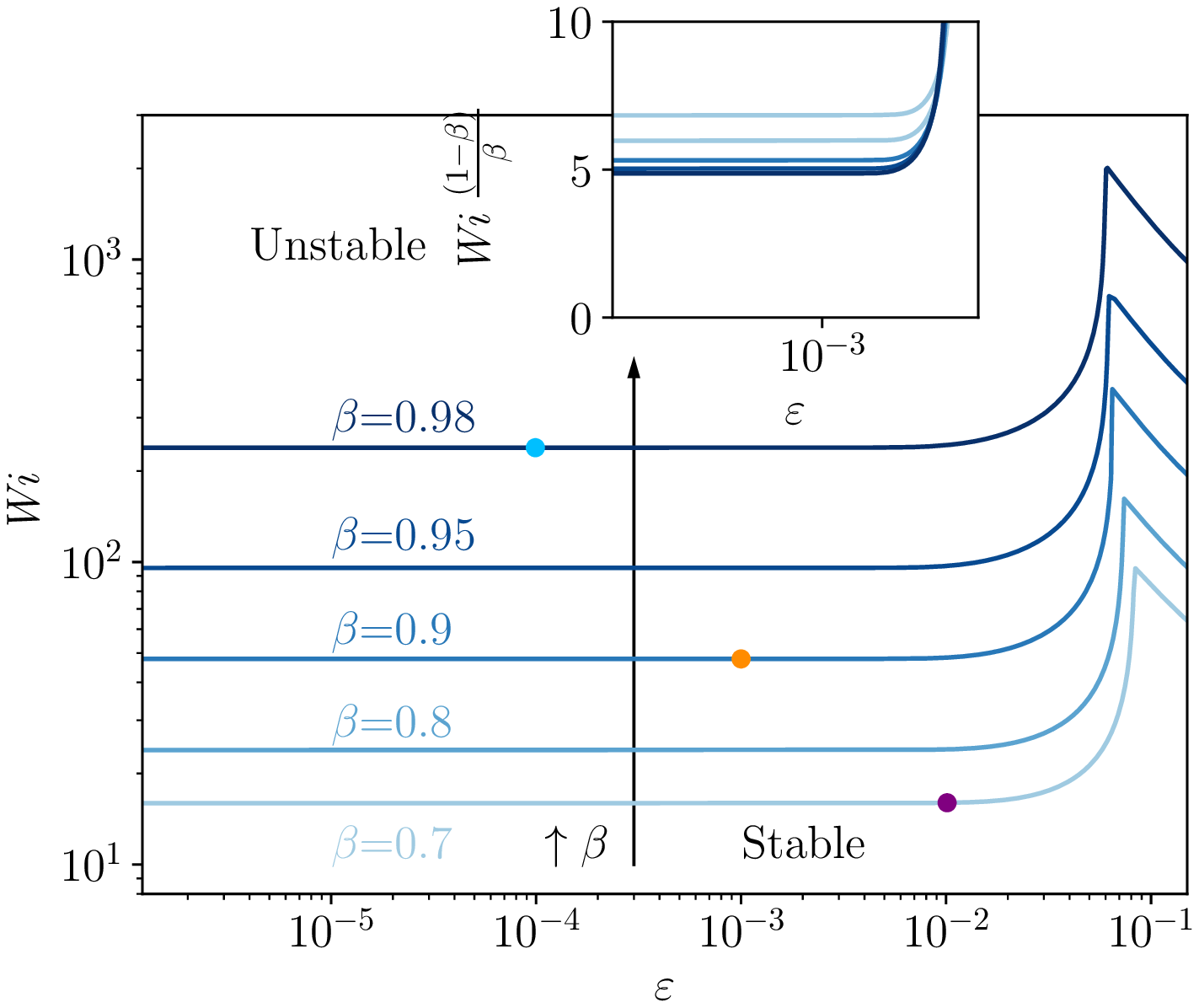} \put(-225,155){$(b)$}\\
    \includegraphics[width=0.89\columnwidth]{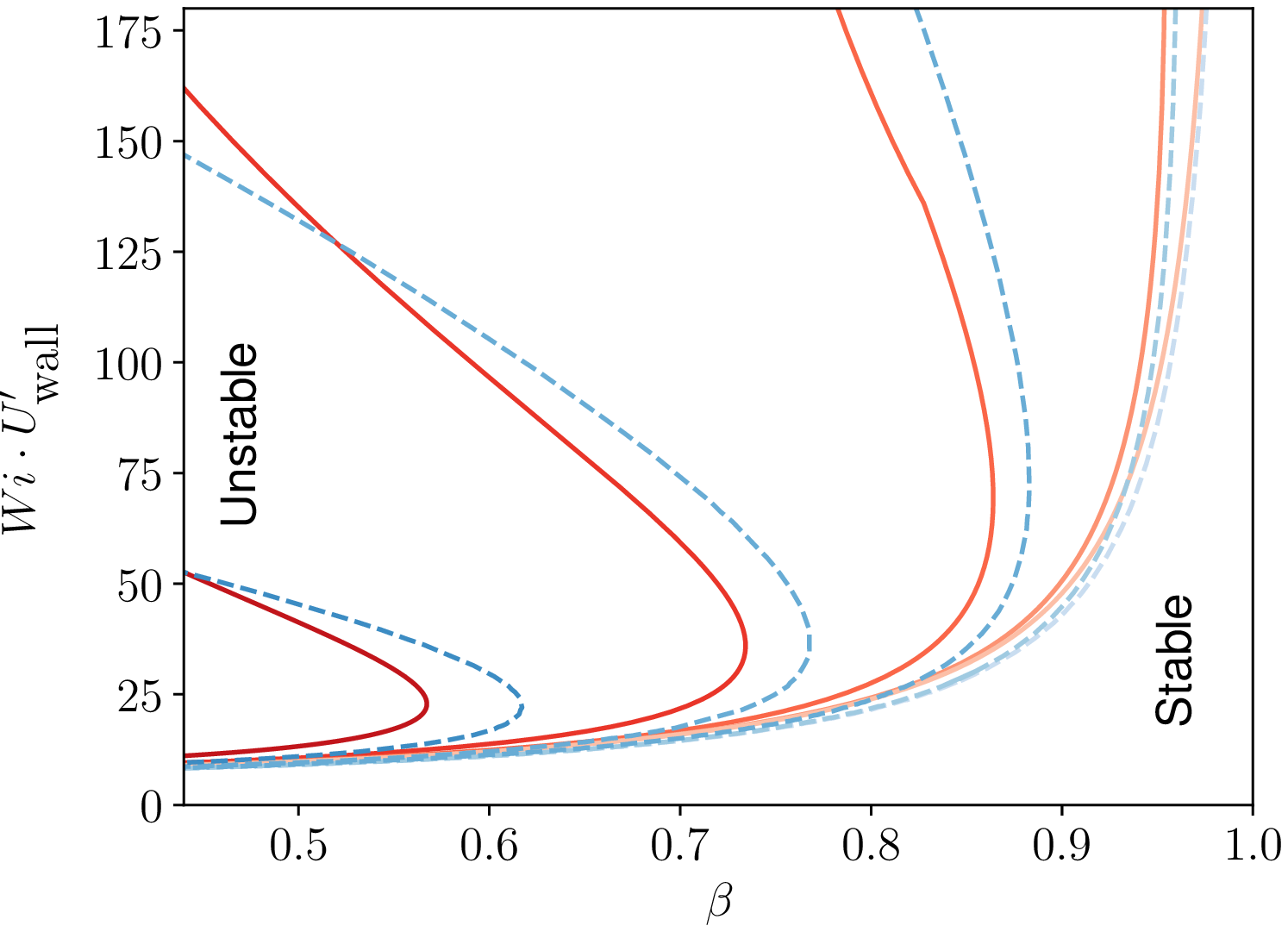}
    \put(-225,155){$(c)$}
\caption{(a) Neutral curves for Oldroyd-B in the $\varepsilon$-$\textit{Wi}$ space for $\beta=0.9$ with different boundary conditions: (i) $\varepsilon=0$ at  the walls (blue), (ii) $\varepsilon\to 0$ only for $\partial_{yy} \mathbf{C}$ (red) and (iii) $\partial_{y}\mathbf{C}=0$ (purple). Dashed lines show the neutral curves for individual wavenumbers $\{1,2,3\}$ with b.c.~(ii). Inset shows the wavenumber $k_x$ along the neutral curves and its scaling with $\varepsilon^{-1/2}$ for all boundary conditions.
(b) Neutral curves in the $\varepsilon$-$\textit{Wi}$ for different $\beta$ in Oldroyd-B. Inset: collapse of the curves in the ultra-dilute limit $\beta\to 1$, $\textit{Wi}\to\infty$. (c) Effect of the adding finite extensibility to \MB{plane Couette flow (red \RK{solid} lines) and plane Poiseuille flow (blue \RK{dashed} lines)} in the FENE-P model. Curves show different values of \MB{$L_{\text{max}}=\{\infty, 600, 200, 100, 60\}$, curves from right  (lighter shade) to left (darker shade)} for $\varepsilon=10^{-3}$.
}
\label{fig:fig2}
\end{figure}
%
%
\begin{figure}
    \includegraphics[width=0.8\columnwidth]{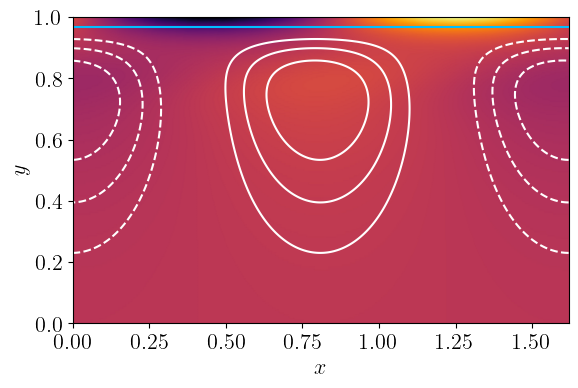} \put(-210,115){$(a)$}\\
    \includegraphics[width=0.75\columnwidth]{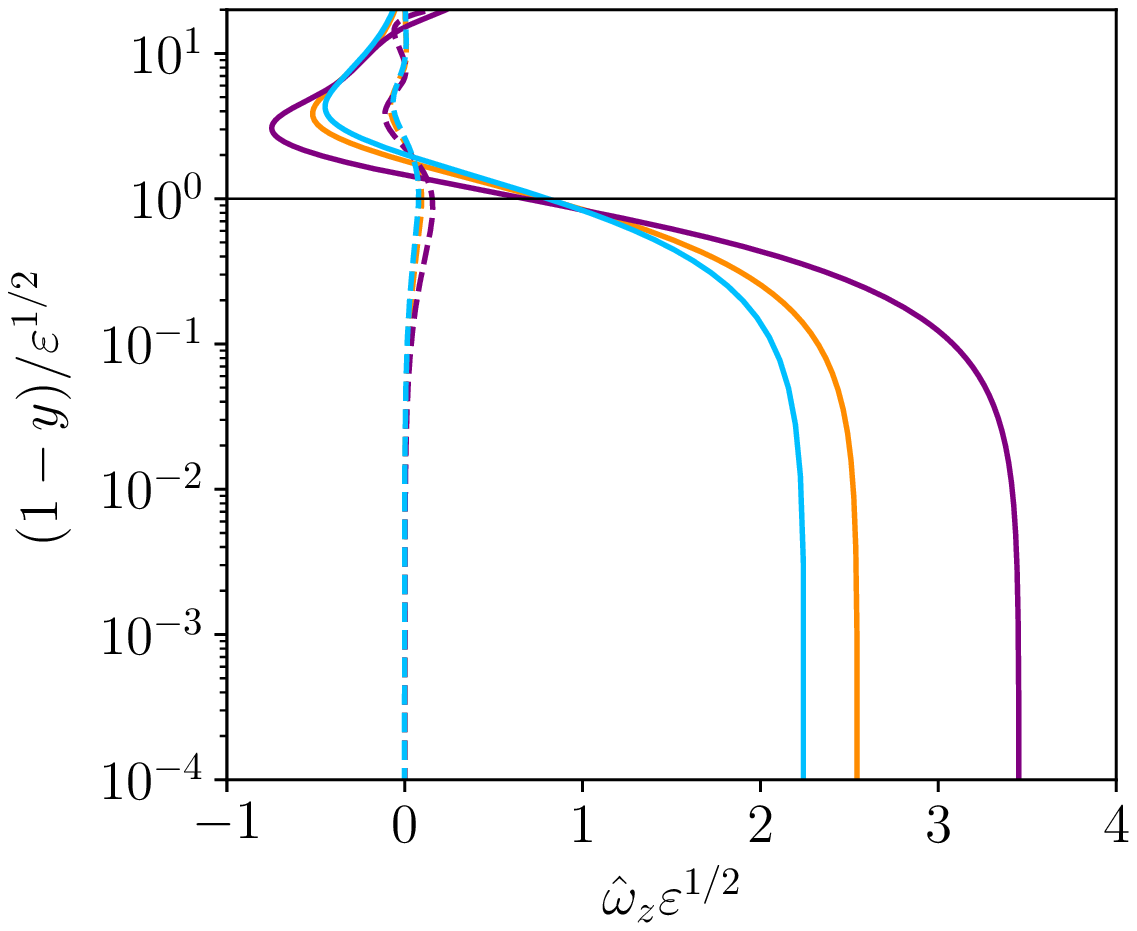}
    \put(-205,130){$(b)$}
\caption{(a) Pseudocolor corresponding to $\text{tr}(\mathbf{C})$ of the upper wall eigenfunctions on the neutral curves for the orange point indicated in figure \ref{fig:fig2}(b). Contour lines indicate the streamfunction (flow left to right). (b) Spanwise vorticity for the corresponding coloured dots in figure \ref{fig:fig2}(b). The eigenfunctions are normalised so that $\|v_{\text{max}}\|=1$.} \label{fig:fig3}
\end{figure}

We map out the unstable region for various parameters and boundary conditions in figure \ref{fig:fig2}(a). In the top left panel of figure \ref{fig:fig2}(a) we observe that the instability persists as $\varepsilon \to 0$ i.e. this is a singular limit for all three boundary conditions and occurs at a constant value of $Wi$ for diffusivities $\varepsilon \leq 10^{-2}$, requiring an increasingly large streamwise wavenumber $k_x \propto \varepsilon^{-1/2}/8$ (\cite{Kumaran_09} did not consider these large wavenumbers and so failed to find instability). 
The imaginary part of the wavespeed \JP{scales with the square root of the diffusivity}, $c_i \propto \varepsilon^{1/2}$, so that the growth rate of the instability, $k_x c_i$, remains $O(1)$ as $\varepsilon \rightarrow 0$.
The unstable region appears unbounded as $Wi \to \infty$ for boundary conditions (i) and (ii) but stability is restored in this limit for (iii). Henceforth boundary condition (i) is used.

In figure \ref{fig:fig2}(b)-(c) we also examine the effect of the viscosity ratio and finite extensibility on the diffusive instability. The instability is realised for decreasing values of $Wi$ at fixed $\varepsilon$ as $\beta$ is reduced (i.e. increasing polymer concentration), and the marginal stability curves collapse when plotted against $Wi(1-\beta)/\beta$ in the ultradilute limit $\beta \to 1$, which is the magnitude of the perturbation stresses $\tau'_{xx}$ and $\tau'_{xy}$ relative to the diffusion terms in the momentum equation. Furthermore, the instability survives for realistic values of the polymer extensibility of $L_{max}=O(100)$, in the FENE-P model, though it is pushed to increasingly low values of $\beta$ and suppressed beyond a critical $Wi$. \MB{Figure \ref{fig:fig2}(c) shows that this instability is also present in plane Poiseuille flow driven by a non-dimensional pressure gradient $\partial_x P=-2$, }
\JP{the neutral curves nearly overlapping when $Wi$ is rescaled by the shear rate at the wall, $U'_{\textrm{wall}}$. 
The quantitative differences observed between the two flow configurations for the specific case considered in \ref{fig:fig2}(c) arise due to the relatively large value of $\varepsilon = 10^{-3}$ (the boundary layer has thickness $\varepsilon^{1/2}$), so that the instability is affected by the non-monotonic channel profile.}
A neutral eigenfunction is shown in figure \ref{fig:fig3}(a), where we visualise contours of the perturbation trace of the polymer conformation and streamlines for an example set of parameters $Wi=47.84,\ \beta=0.9,\ \varepsilon=10^{-3}$. The spanwise vorticity is shown below in figure \ref{fig:fig3}(b) for various values of $\varepsilon$, $\beta$ and $\textit{Wi}$.

%
%
\begin{figure*}
    \centering
    \begin{minipage}{0.60\textwidth}
    \includegraphics[width=1.0\textwidth]{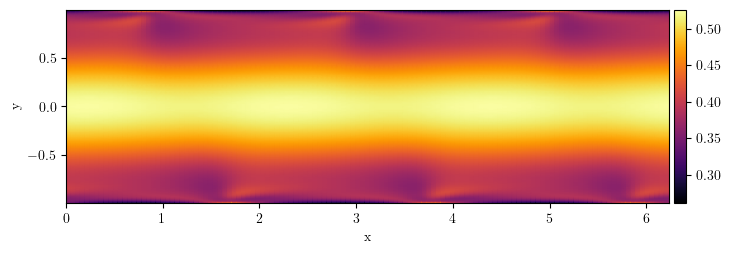} \put(-310,100){$(a)$}\\
    \includegraphics[width=1.0\textwidth]{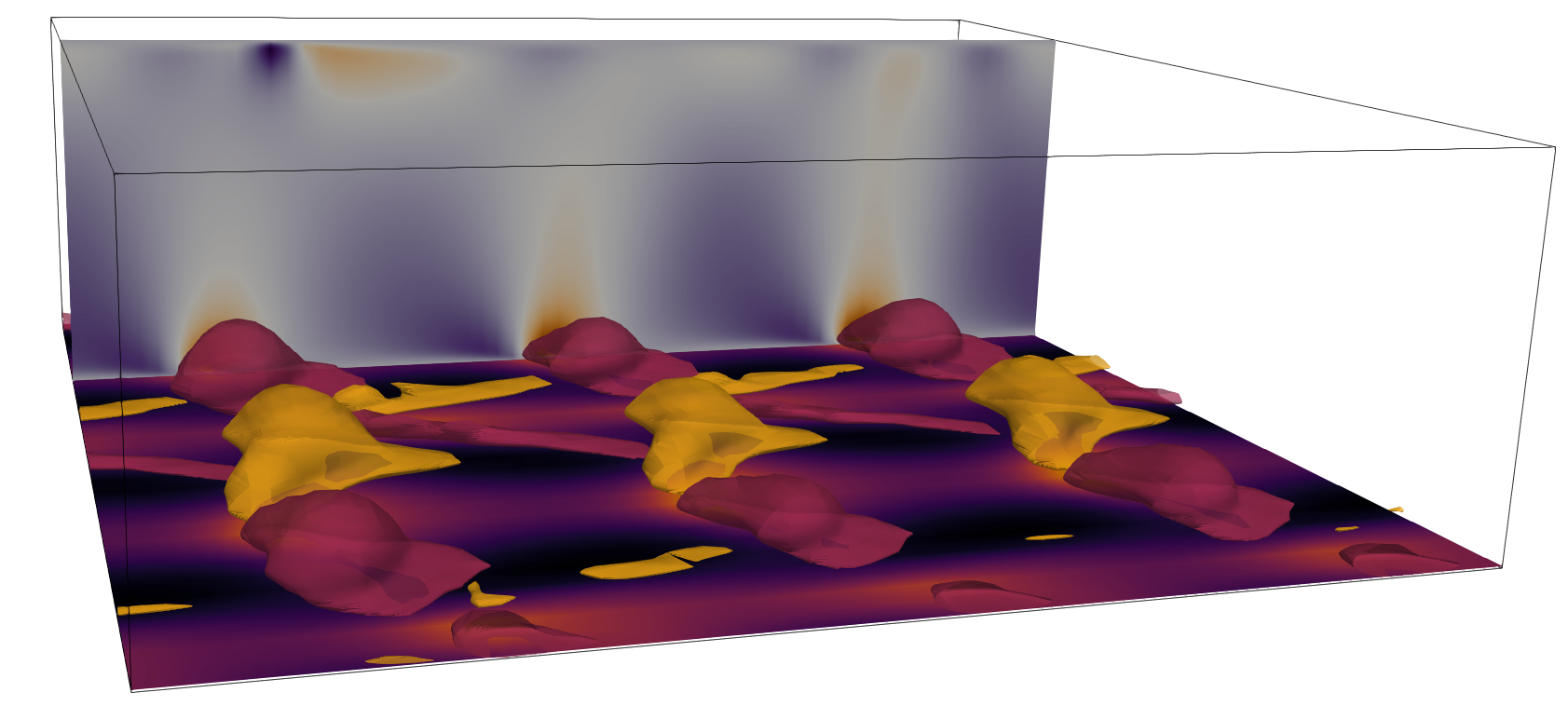} \put(-310,130){$(c)$}
    \end{minipage}
    \begin{minipage}{0.32\textwidth}
    \includegraphics[width=1.0\textwidth]{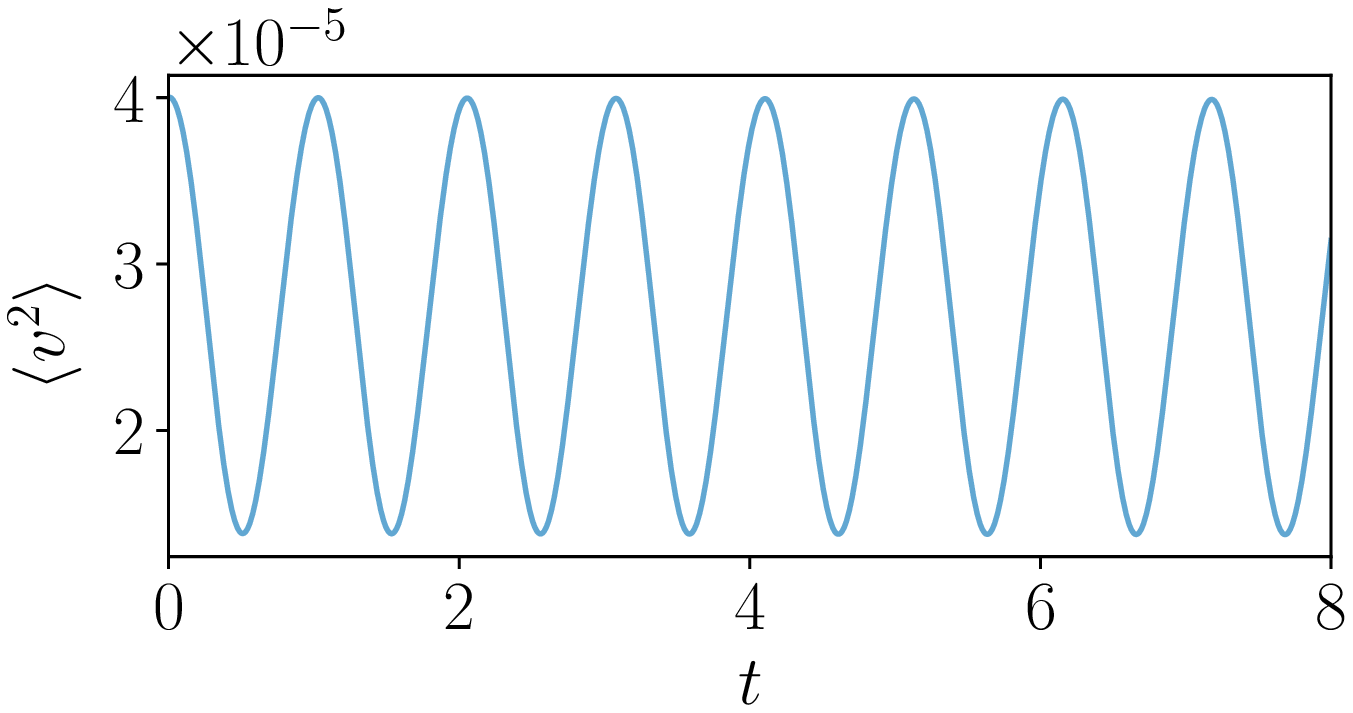} \put(-165,80){$(b)$}\\
    \includegraphics[width=1.0\textwidth]{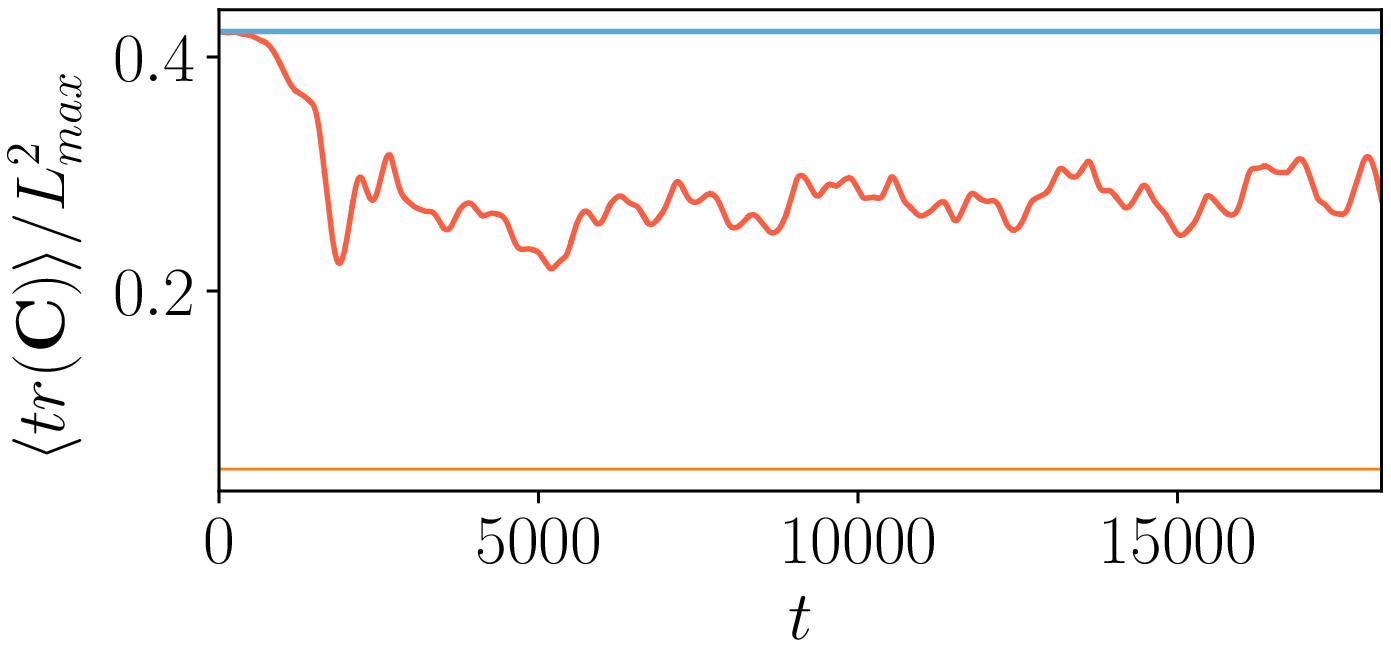} \put(-165,80){$(d)$}\\
    \includegraphics[width=0.9\textwidth]{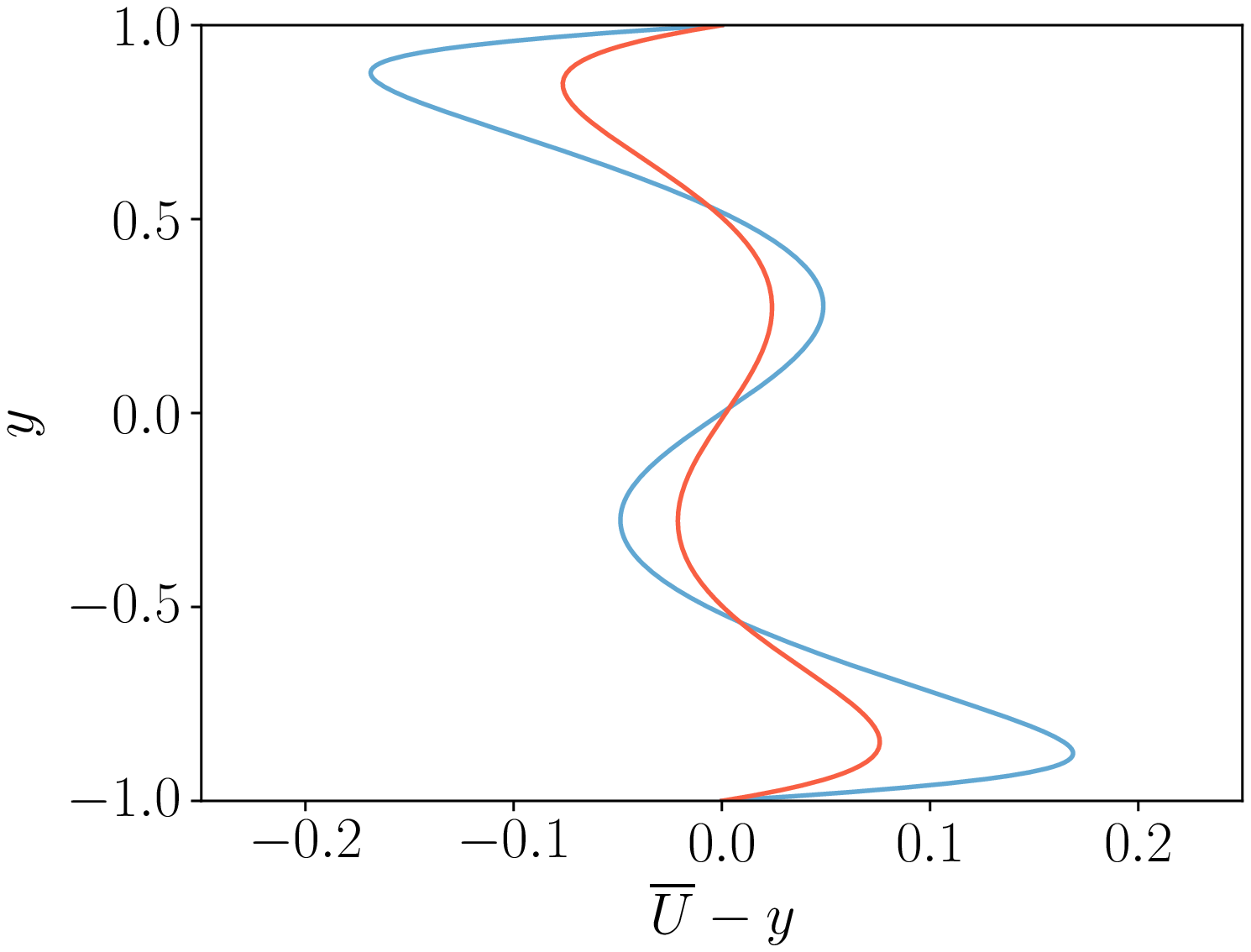}
    \put(-154,105){$(e)$}
    \end{minipage}
    \includegraphics[width=0.288\textwidth]{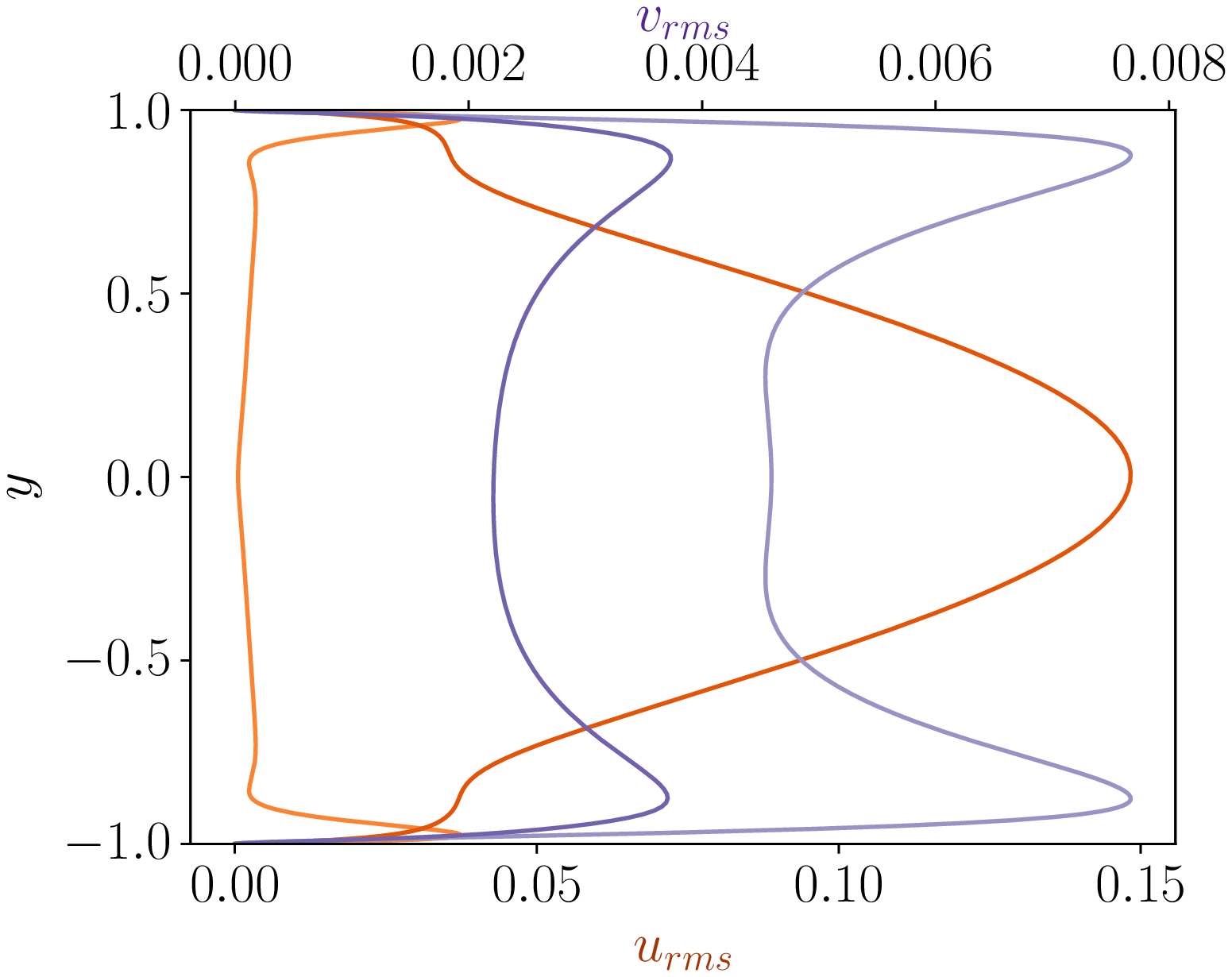}\put(-151,105){$(f)$} \hspace{1.0mm}
    \includegraphics[width=0.288\textwidth]{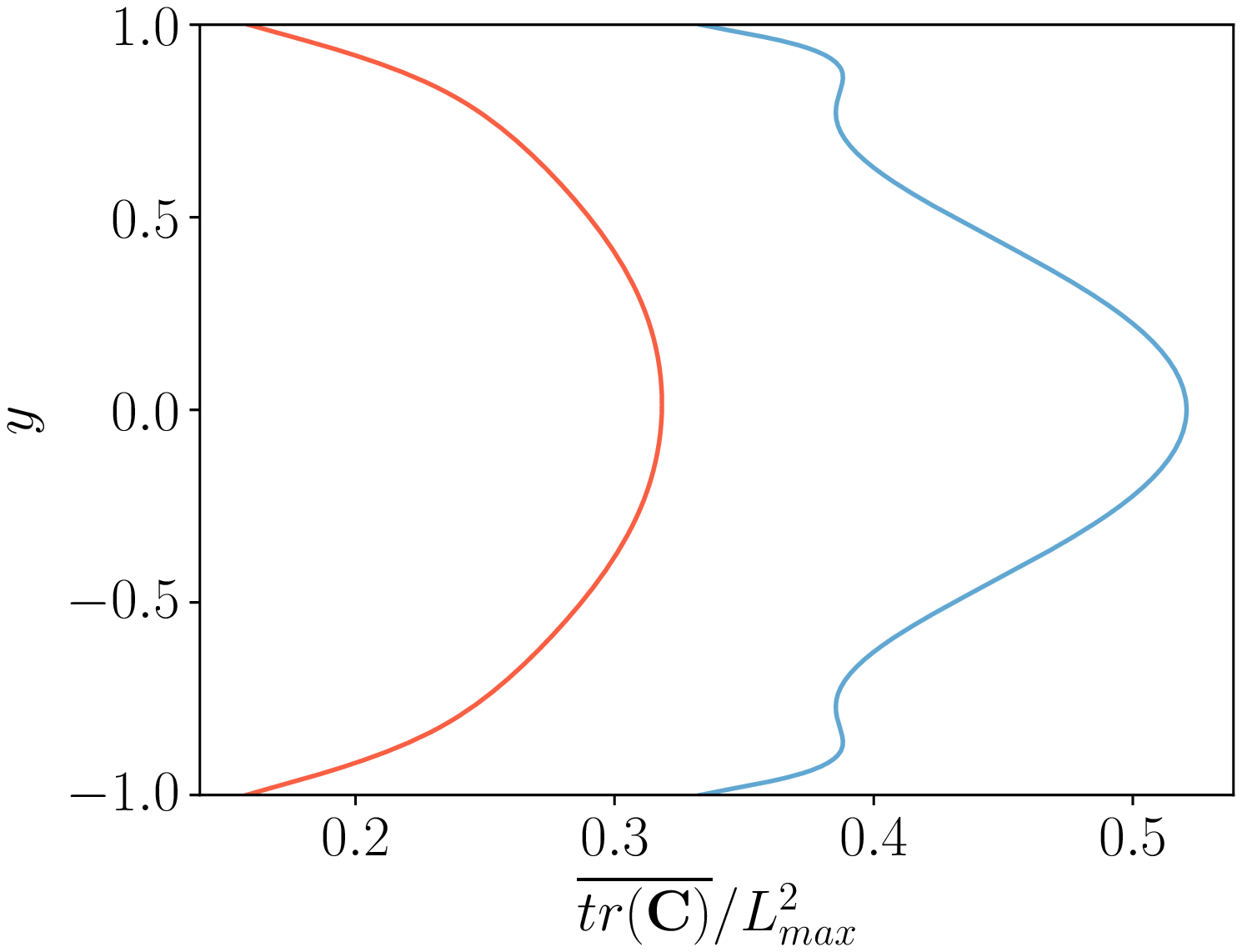} \put(-151,105){$(g)$} \hspace{1.0mm}
    \includegraphics[width=0.288\textwidth]{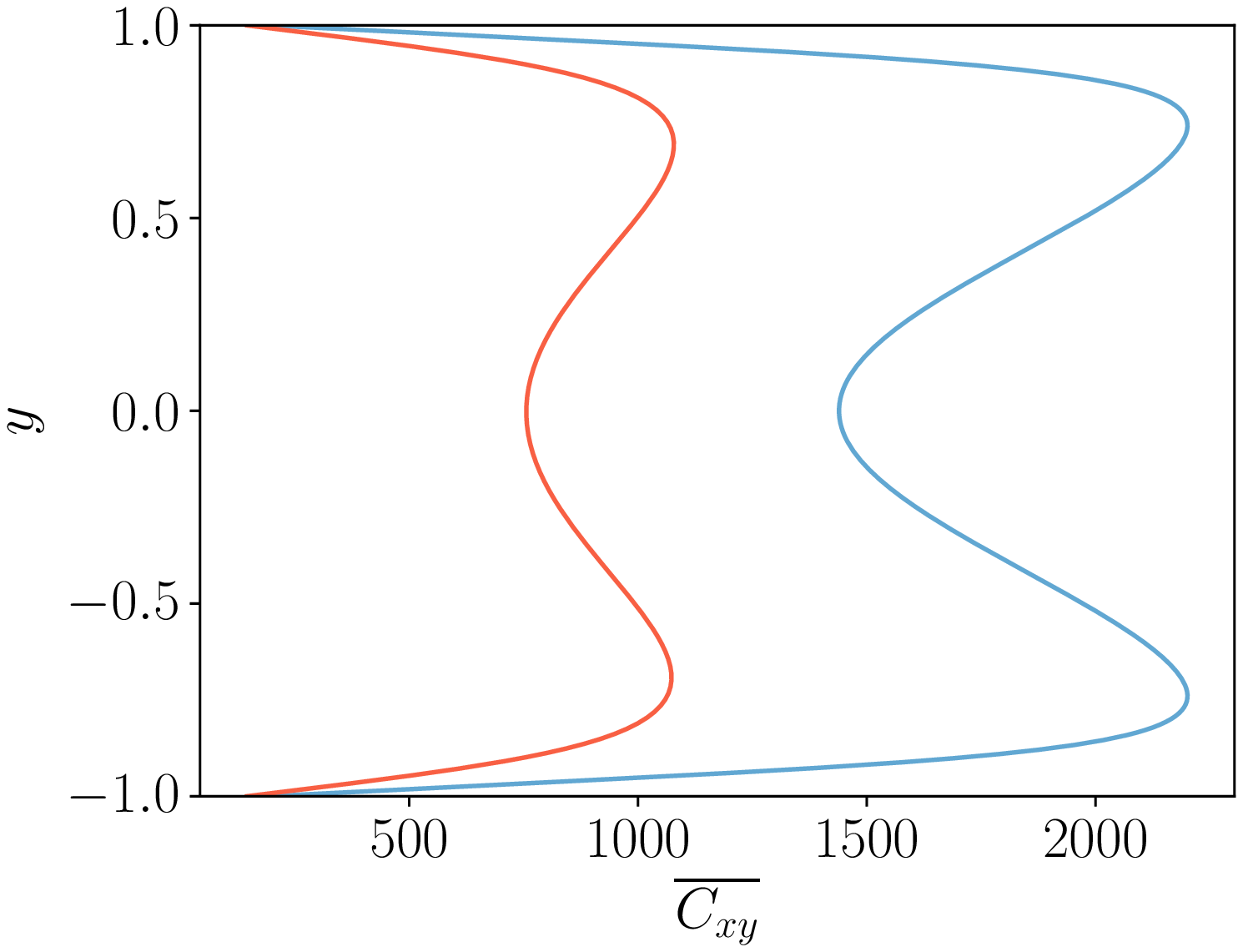} \put(-151,105){$(h)$}
    \caption{Direct numerical simulations of inertialess viscoelastic plane Couette flow for $\textit{Wi}=100,\ \beta=0.9,
    \ L_{max}=600,\ \varepsilon=10^{-3}$. (a) Instanteneous visualization of $\langle \text{tr} (\mathbf{C})\rangle$ of the two-dimensional periodic orbit identified. (c) Instanteneous visualization of the chaotic three-dimensional state identified. The horizontal slice is a pseudocoulor of $\text{tr}(\mathbf{C})$ at $y=-0.99$, the vertical slice is a pseudocoulor of vertical velocity, $v$. The contour indicate positive and negative values of the spanwise velocity, $w$. Flow from left to right. (b) Time series of $\langle v^2\rangle$ for the two-dimensional nonlinearly saturated state corresponding to a low-amplitude high-frequency periodic orbit. (d) time series of the volume average $\text{tr}(\mathbf{C})$ for the chaotic three-dimensional flow with the parameters above (red), the two-dimensional nonlinearly saturated state (blue), and the laminar state (orange). (e)-(g) and (h) time average mean quantities -- denoted $(\cdot)$ -- for the two-dimensional limit cycle and the three-dimensional chaotic state; (e) deviation of the mean velocity from $U=y$ (2D blue, 3D red). (f) $u_{\text{rms}}$ (purple: 2D light, 3D dark) and $v_{\text{rms}}$ (orange: 2D light, 3D dark) (g) $\overline{ \text{tr}(\mathbf{C})}/L_{\text{max}}$ (2D blue, 3D red). (h): $\overline{C_{xy}}$ (2D blue, 3D red).}
    \label{fig:DNS}
\end{figure*}
%
%
\MB{The mechanism of instability is reminiscent of the destabilizing effect of viscosity in Newtonian channel flow which subtly adjusts the relative phases of key dynamical \cite{lin1955theory}. }
\JP{In the polymeric diffusive instability} perturbations to the polymer field are highly stretched in a boundary layer of size $O(\varepsilon^{1 / 2})$ at one of the walls where the perturbation vorticity is concentrated. 
Without stress diffusion, the phase of the ensuing polymer stress is poorly matched to that of the velocity field so there is not enough positive feedback \JP{in the momentum equation}. 
A diffusive phase lag in the polymer stretch response, however, allows the polymer stress to sufficiently reinforce the flow which created it, thereby producing an instability (see Supplementary Material for a more detailed description).

The instability described here is realised \MB{for commonly used parameters in numerical simulations of dilute polymer solutions} (e.g. $Wi \sim 50$ at $\beta = 0.9$) over a wide range of realistic diffusivities. For example, in a microfluidic context, $\varepsilon$ can range from $O(10^{-6})$ for long polymer molecules formed by random chains with contour length much larger than the persistence length to $O(10^{-3})$ for short polymers.
The instability exists in a boundary layer of size $\approx 10\sqrt{\varepsilon}$ (see figure \ref{fig:fig3}(b)) and has an optimal wavenumber $k_x \approx 0.125/\sqrt{\varepsilon}$  consistent with this (see figure 1(a) in Supplementary Material). Dimensionally, this is a lengthscale of $10\sqrt{\varepsilon} \,h= 10 R/\sqrt{Wi} \sqrt{ k_B T \lambda/6 \pi \mu_s R^3} $ using the Stokes-Einstein relation ($k_B T$ is the thermal energy and $R$ is the polymer gyration radius) and with typical values (e.g. see \cite{Burshtein17}) reduces to $\approx 10R/\sqrt{Wi}$ consistent with the simple scaling relationship $D \sim R^2/\lambda$. The optimal instability lengthscale therefore approaches the polymer lengthscale for $Wi \gtrsim 100$ although larger-scale instability modes, which offer a better scale separation, also exist away from the neutral curve (see figure 1(b) in the Supplementary Material).

%
%

We now show using the FENE-P model that this instability forms a pathway to chaos in planar Couette flow.
The nonlinear evolution of the diffusive instability is examined  by first conducting two-dimensional DNS of the governing equations 
(\ref{eqn:governing}) 
using the open-source codebase Dedalus \cite{burns2020dedalus}. We perform calculations at parameter settings $Wi=100$, $\beta = 0.9$, $L_{max}=600$ and $\varepsilon=10^{-3}$ in a box of length of length $L_x=2\pi$ with resolution $[N_x,N_y]=[128,256]$. The simulations are initialised with low amplitude white noise, which excites the most unstable mode $k_x=3$. The unstable mode amplifies exponentially before saturating onto a limit cycle with period $T \sim h/U_0$.
We report a snapshot of the instantaneous polymer stretch field in figure \ref{fig:DNS}(a) alongside the volume-averaged -- denoted $\langle \,\cdot\, \rangle$ -- vertical velocity squared in figure \ref{fig:DNS}(b). While the polymer is significantly stretched relative to the laminar value, the velocity fluctuations are relatively low amplitude, $v = O(10^{-2}U_0)$. 

In flows with inertia, EIT can be realised in purely two-dimensional configurations \cite{sid2018two}, which is not something we observe in our inertialess calculations. The stability of the limit cycle is examined in a three-dimensional configuration with spanwise width $L_z = 2\pi$ by adding small-amplitude white noise in the velocity component $w$. Three-dimensional simulations were performed with resolutions 
ranging from $[N_x, N_y,N_z]=[64,128,32]$ to $[N_x, N_y,N_z]=[128,256,64]$ to check robustness. A time series of the volume-averaged polymer trace from this computation is reported in figure \ref{fig:DNS}(d) which shows a departure from the simple time-periodic solution to a chaotic trajectory which is maintained for over a $10^4 h/U_0$ time period. We believe this to be a numerical realisation of elastic turbulence in parallel flow. 

An instantaneous snapshot of the chaotic flow can be seen in figure \ref{fig:DNS}(c). The horizontal pseudocolor plane located at $y=-0.99h$ (in the boundary layer at the lower wall for $\varepsilon =10^{-3}$) shows how $tr(\mathbf{C})$ is modulated in the both the streamwise and spanwise direction forming high and low high stretch regions. The vertical back plane contours the vertical velocity $v$, which is amplified in small scale near-wall patches reminiscent of suction and ejection events in high Reynolds number wall bounded turbulence. The three-dimensional contours (in the box) represent $w$, illustrating the spanwise motion of the flow. 

The statistics of both the limit cycle and 
the three-dimensional chaotic attractor
are examined further in panels (e)-(h) of figure \ref{fig:DNS}, which indicates (i) that the limit cycle is a significant departure from the laminar base state, (ii) the chaos departs from the limit cycle while retaining certain features (iii) the polymer is substantially stretched in the centre of the domain for both the limit cycle and in the chaotic flow. There are also sharp variations of $u_{rms}$ in a boundary layer $\delta \propto \varepsilon^{1 / 2}$ at the walls (verified in computations at various $\varepsilon$ but not shown). The values of $u_{rms}$ in ET are significantly different from the limit cycle and reach their largest magnitude at the centreline.

%
%
The presence of polymeric stress diffusivity is commonly disregarded in the linear stability analyses of viscoelastic flows \cite{larson1990purely,pan2013nonlinear,garg2018viscoelastic,khalid2021centre,sanchez22, van2022characterizing}.
However, we have found that its presence, even at vanishingly small values in the FENE-P model, fundamentally changes the stability of viscoelastic plane Couette flow in the absence of inertia (i.e. vanishing polymer stress diffusion is a singular limit). 
This polymeric diffusive instability (PDI) is a `wall' mode, travelling with  roughly the wall speed, and has a streamwise wavelength comparable to the boundary layer thickness. This last feature could easily have led to the instability being dismissed in the past as a numerical instability. The onset of instability is found at $Wi\approx 8$ independently of $L_{max}$ and $\varepsilon$ for a broad range of $\beta$. DNS of the instability leads to a stable periodic orbit in 2D and to spatiotemporal chaos in 3D, the first numerical realisation of such \JP{a flow state} in a \MB{wall-bounded} parallel flow configuration. 
\RK{This chaotic state is reproducible in other viscoelastic shear flows with and without inertia \footnote{manuscript in preparation}}.

The fact that the PDI operates on a scale approaching the polymer gyration radius at higher $Wi$ ($\gtrsim 100$) brings into question whether (a) FENE-P remains a good viscoelastic model there, and (b) whether this instability could be experimentally observed. Certainly at lower $Wi=O(10)$ and for less optimal (larger scale) PDI modes, there should be sufficient scale separation for the continuum approximation. The bigger issue is whether FENE-P with polymer stress diffusion incorporates enough physics to be realistic when the PDI drives the dynamics.  Preliminary calculations in the two-fluid model of \cite{Mavrantzas92}(simplified to a solution of Hookean dumbells \cite{Mavrantzas94, black2001slip}, see Supplementary Material) suggest that the added presence  of polymer concentration  diffusion stabilises the PDI.  This would suggest that the PDI will be not be seen in the laboratory. If that is the case, our results then call into question the continuing use of the FENE-P model which can only be numerically integrated forward in time if polymer stress diffusion is present (either artificial or numerical).

\vspace{1cm}
The authors gratefully acknowledge the support of EPSRC through grant EP/V027247/1. The authors declare no competition of interest.

\newpage

\preprint{APS/123-QED}

\title{Supplementary Material\\Polymeric diffusive instability leading to elastic turbulence in plane Couette flow}

\author{Miguel Beneitez}
\affiliation{
 DAMTP, Centre for Mathematical Sciences, Wilberforce Road, Cambridge CB3 0WA, UK
} 
 
\author{Jacob Page}%
\affiliation{
 School of Mathematics, University of Edinburgh, EH9 3FD, UK
}

\author{Rich R. Kerswell}
\affiliation{
 DAMTP, Centre for Mathematical Sciences, Wilberforce Road, Cambridge CB3 0WA, UK
}

\date{\today}
\maketitle

%
%
%
 
\section*{Supplementary material}\label{sec11}

\subsection{Linear stability analysis} 

Adding a small perturbation, $\phi^*$, to the base state $\Phi$ so that the total  field is $\phi=\Phi+\phi^{*}$, the linearised equations derived from (1a)-(1c) in the main manuscript are

\begin{widetext}
\begin{subequations}
\begin{align}
    \partial_x p^* &= \beta\left(\partial_{xx}u^* + \partial_{yy}u^* \right) + (1-\beta)\left(\partial_x \tau^*_{xx} + \partial_{y}\tau^*_{xy} \right),\\
    \partial_y p^* &= \beta\left(\partial_{xx}v^* + \partial_{yy}v^* \right) + (1-\beta)\left(\partial_x \tau^*_{xy} + \partial_{y}\tau^*_{yy} \right),\\
       \partial_t c^*_{xx}+U\partial_x c^*_{xx}+C_{xx}' v^{*} +\tau^*_{xx}&=2\left(C_{xx}\partial_x u^*+ U' c^*_{xy} + C_{xy}\partial_y u^{*}\right) +  \varepsilon\left(\partial_{xx}c^*_{xx}+\partial_{yy}c^*_{xx}\right), \\
    \partial_t c^*_{yy}+U\partial_x c^*_{yy}+C_{yy}' v^{*} +\tau^*_{yy}&=2\left(C_{xy}\partial_xv^{*} + C_{yy}\partial_y v^{*}\right)  \hspace{1.7cm}+ \varepsilon\left(\partial_{xx}c^*_{yy}+\partial_{yy}c^*_{yy}\right),\\
    \partial_t c^*_{zz}+U\partial_x c^*_{zz}+C_{zz}' v^{*} +\tau^*_{zz}&= \hspace{5.25cm}\varepsilon\left(\partial_{xx}c^*_{zz}+\partial_{yy}c^*_{zz}\right),\\
        \partial_t c^*_{xy}+U\partial_x c^*_{xy}+C_{xy}' v^{*} +\tau^*_{xy}&=\left(C_{xx}\partial_x v^{*}+U'c^*_{yy}+C_{yy}\partial_y u^{*}\right) \hspace{0.15cm}+ \varepsilon\left(\partial_{xx}c^*_{xy}+\partial_{yy}c^*_{xy}\right),
    \\
    \partial_x u^{*}+\partial_y v^{*} & = 0,
\end{align}
\end{subequations}
where 
\begin{subequations}
\begin{align}
    \tau^*_{ij} &= \frac{1}{Wi}\left(\text{tr}(\mathbf{c}^{*})\frac{f^2(\text{tr}\ \mathbf{C})}{L_{\text{max}}^2}C_{ij}+f(\text{tr}\ \mathbf{C}) c^*_{ij}\right),\\
    \partial_x \tau_{ij}^* &=\frac{1}{Wi}\left(\partial_x\text{tr}\ \mathbf{c}\frac{f^2(\text{tr}\ \mathbf{C})}{L_{\text{max}}^2}C_{ij}+f(\text{tr}\ \mathbf{C})\partial_x c^*_{ij}\right),\\
        \partial_y \tau_{ij}^*& =\frac{1}{Wi}\left(\partial_y\text{tr}\ \mathbf{c}\frac{f^2(\text{tr}\ \mathbf{C})}{L_{\text{max}}^2}C_{ij} + \text{tr}\ \mathbf{C}\partial_y \left[\frac{f^2(\text{tr}\ \mathbf{C})}{L_{\text{max}}^2}C_{ij}\right] + 
          c^*_{ij}\partial_y f(\text{tr}\ \mathbf{C}) +  f(\text{tr}\ \mathbf{C})\partial_y c^*_{ij}\right).
\end{align}
\end{subequations}
\end{widetext}
(primed quantities denote base flow derivatives with respect to the wall-normal direction and $C'_{xx}=C'_{xy}=C'_{yy}=C'_{zz}=0$ for plane Couette flow).
We consider the exponential growth of perturbations and apply a Fourier transform in the homogeneous $x$ direction  by introducing the ansatz $\phi^* = \hat{\phi}(y)e^{i k_x(x-c t)}$, where $k_x$ is the the streamwise wavenumber and $c=c_r+ic_i$ the complex phase speed of the perturbations, with $c_i>0$ for linear instability.

\begin{figure}
     \includegraphics[width=0.89\columnwidth]{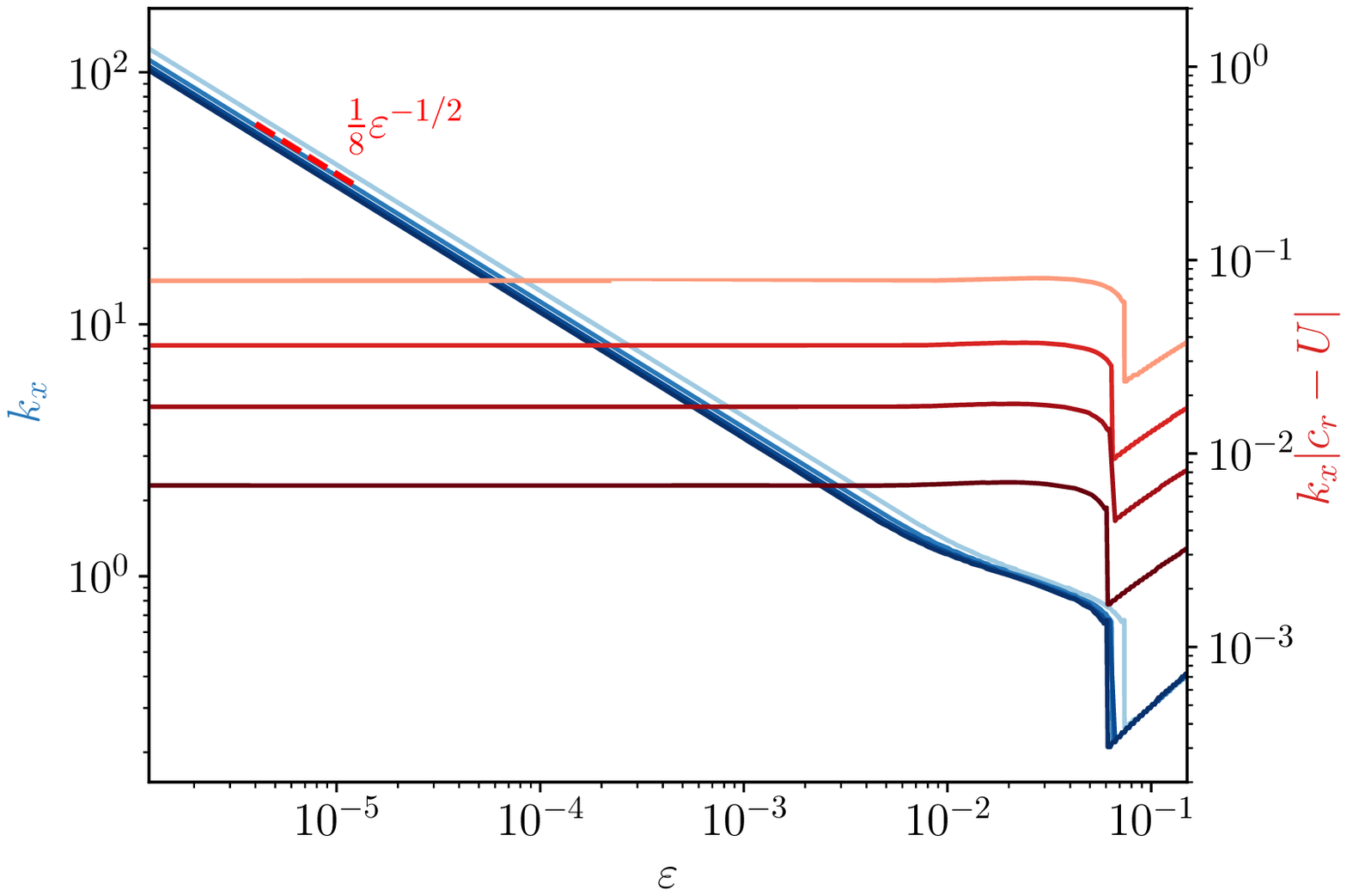}\put(-225,155){$(a)$} \\
     \includegraphics[width=0.80\columnwidth]{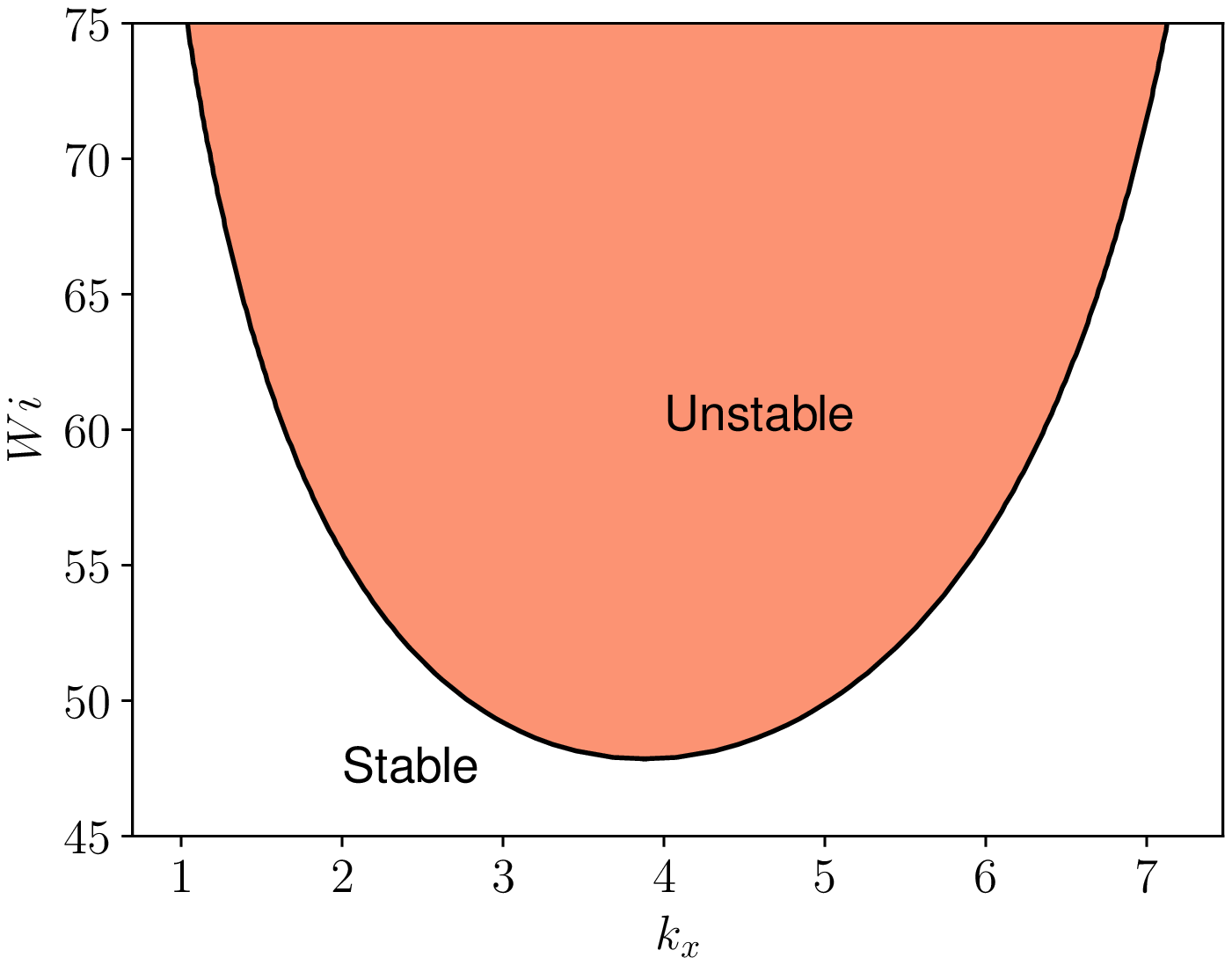}\put(-225,155){$(b)$}
\caption{(a) Wavenumbers $k_x$ (in blue) corresponding to PDI for Oldroyd-B along the neutral curves in Fig. 2(a) for the same values of $\beta$. Dashed red line shows that $k_x\approx 1/8\varepsilon^{-1/2}$ for the critical wavenumber of the instability. Solid red shaded curves show $k_x|c_{r}-U|$, of the PDI along the same neutral curves. (b) Unstable wavenumbers for PDI for $\beta=0.9$, $\varepsilon=10^{-3}$. The shaded region shows how the range of unstable $k_x$ increases beyond criticality.
}
\label{fig:SFig}
\end{figure}

Figure \ref{fig:SFig} shows that the optimum wavenumber for the PDI scales as $k_x \approx 0.125/\sqrt{\varepsilon}$ and that $1-c_r=O(\sqrt{\varepsilon})$  for the PDI localised at the boundary moving at $1$ on the neutral curve. This is consistent with the scaling of the growth rate $k_x c_i=O(1)$  as $\varepsilon \rightarrow 0$ away from the neutral curve.

\subsection{Instability mechanism}

For an Oldroyd-B fluid, the instability persists in the ultra-dilute limit of  $Wi \rightarrow \infty$  and $\beta \rightarrow 1$ with  $\Lambda=Wi(1-\beta)/\beta$ staying finite: see figure 2 in the main text.
%
The leading order equations for the perturbation in this limit are
\begin{subequations}
\begin{align}
i k_x p^* -\left(\partial^2_{y}-k_x^2 \right) u^* &= i k_x \tau_{xx}+\underbrace{D \tau_{xy}}, \label{m1}\\
\partial_y p^* -\left(\partial^2_{y}-k_x^2 \right) v^* &=\underbrace{i k_x \tau_{xy}}, \label{m2} \\
 \left[ik_x(U-c)-\varepsilon\left(\partial^2_{y}-k_x^2\right) \right]\tau_{xx} &=\underbrace{4ik_x \Lambda u^*} +2U' \tau_{xy},\label{m3} \\
 \left[ik_x(U-c) -\varepsilon\left(\partial^2_{y}-k_x^2\right) \right]\tau_{xy} &= 2ik_x \Lambda v^*,   \label{m4}  \\
 ik_x u^{*}+\partial_y v^{*} & = 0. \label{m5}
\end{align}
\end{subequations}

%
where perturbation variables are taken proportional to $\exp(ik_x(x-ct))$ and $\tau_{ij}:=(1-\beta)\tau^*_{ij}$ is a rescaled stress. Note the appearance of $\Lambda$ in the polymer conformation evolution equations (\ref{m3} and \ref{m4}) -- this is the rescaled base normal stress, $T_{xx}$.
We have performed numerical experiments that indicate that the terms in equations (\ref{m1}--\ref{m5}) highlighted with underbraces can be dropped without suppressing the instability. 
We can therefore describe the instability mechanism as follows:
In equation (\ref{m4}) the highly-tensioned base state polymers are tilted by $\partial_x v^*$,  to generate a perturbation shear stress $\tau_{xy}$. 
This perturbation shear stress is then deformed by the base shear to drive a streamwise stress perturbation, $\tau_{xx}$, in equation (\ref{m3}) which can then enhance the streamwise velocity in the momentum equation (\ref{m1}), provided there is a phase adjustment through polymer stress diffusion.
The phase adjustment through the diffusion terms in (\ref{m3}) and (\ref{m4}) is crucial for the positive feedback as the reduced system is stable without it.
For example, at $\Lambda=3$, $k_x=1.4$ and $\varepsilon=0.001$, using Neumann boundary conditions on the stress for simplicity, we find PDI with $c_i=0.0024$, which vanishes on setting $\varepsilon = 0$.

\subsection{Numerical method} 

%
%

The direct numerical simulations (DNS) of eqs. (1a)-(1c) in the main text were performed using the numerical codebase Dedalus \cite{burns2020dedalus} (the Dedalus codebase has been extensively used in flow simulation \cite{burns2020dedalus,dubief2022first,buza2022finite}). We consider a computational domain of size $[L_x,L_y,L_z]=[2\pi,2,2\pi]h$. The quantities $\mathbf{C}$ and $\mathbf{u}$ are expanded in $N_x$ and $N_z$ Fourier modes in the streamwise ($x$) and spanwise ($z$) directions respectively. An expansion in $N_y$ Chebyshev modes is employed in the wall-normal ($y$) direction. Time integration is performed with a 4th-order exponential Runge-Kutta (RK443) scheme with fixed time step $\Delta t=5\times 10^{-3}$. The two-dimensional simulations were performed with a resolution $[N_x,N_y]=[128,256]$ and the three-dimensional ones used $[N_x,N_y,N_z]=[64,128,32]$. The consistency of the results was checked with double the resolution in both cases, obtaining the same dynamical behaviour. The two-dimensional simulations were initialised by adding random noise to the undisturbed laminar state. We tested the three-dimensional stability of the two-dimensional state by adding small amplitude random perturbation in the spanwise component of the velocity $w$. \\

\subsection{Polymer concentration diffusion}

The effect of introducing polymer concentration diffusion was checked by considering the two-fluid model of \cite{Mavrantzas92} simplified to a solution of Hookean dumbells \cite{Mavrantzas94, black2001slip}. Using our non-dimensionalisation, this model is
\begin{align}
\boldsymbol \nabla p = \beta \Delta \mathbf{u} &+(1-\beta) \nabla \cdot \btau, \nonumber \\
\nabla \cdot \mathbf{u} &= 0, \nonumber \\
\partial_t \C + \mathbf{u}\cdot \nabla \C +\btau &= \C \cdot \nabla \mathbf{u}_p + (\nabla \mathbf{u}_p)^{T}\cdot \C + \varepsilon\nabla^{2}\C,   \nonumber \\
\partial_t n + \mathbf{u}\cdot\nabla n &= \varepsilon\nabla \cdot(\nabla n - Wi \nabla \cdot \mathbf{\btau})
\nonumber 
\end{align}
with stress
\begin{equation}
\btau := \frac{1}{Wi}(\C - n\mathbf{I}),  \nonumber 
\end{equation}
where $n$ is the polymer concentration and the definition of the mass-average velocity is

\begin{equation}
    \mathbf{u}_p := \mathbf{u}+\gamma\varepsilon \textit{Wi}\,(-\nabla n +Wi \nabla \cdot \btau).
\end{equation}
We solve the linear stability problem for infinitesimal disturbances superimposed on  the base flow
\begin{equation}
(\mathbf{u_b},\tau_{xx},\tau_{xy},\tau_{yy},n,p)=(\,y\hat{\bf x},2Wi,1,0,1,1) 
\end{equation}
with no-slip boundary conditions on the velocity field and no flux conditions on the stress and the concentration field (the latter given by $\hat{ {\bf n} } \cdot \mathbf{u}_p=0$). PDI is present within this model for $\gamma< 2\cdot 10^{-3}$, however for $\gamma\geq 2\cdot 10^{-3}$ (and $\gamma=1$ in \cite{Mavrantzas92}) no linear instability was found over $\beta \in [0.001 , 0.9998]$ and $Wi \in [4, 1400]$ .

\bibliography{apssamp}

\end{document}